\documentclass[%
preprint,
showpacs,preprintnumbers,
  amsmath,amssymb,
  aps,
 prl
]{revtex4-1}
 
\usepackage{amsfonts}
\usepackage{amssymb}
\usepackage[margin=2cm]{geometry}
\usepackage[dvips]{graphicx}
 
\usepackage[T1]{fontenc}
\usepackage[latin2]{inputenc}
 
\usepackage{amsmath}
\usepackage{bbm}
\usepackage{youngtab}
 
\begin{document}
 
\title{Flux qubits shed a new light on BCS theory and high-$T_c$ superconductivity}

\author{Robert Alicki}

\affiliation{Institute of Theoretical Physics and Astrophysics, University
of Gda\'nsk, Wita Stwosza 57, PL 80-952 Gda\'nsk, Poland}

\date{\today}
\begin{abstract}
A simple microscopic model of a small superconducting loop interrupted by  Josephson junction (flux qubit)
allows to compute from the experimental data of Wal et.al \cite{Wal} an important parameter - the density of Cooper pairs at zero temperature. This density is determined by the cut-off energy in the BCS model and agrees with the original BCS suggestion  but is lower by two orders of magnitude than the value accepted in the modern literature. The immediate consequences of this result are: the validity of the strong coupling BCS model, a plausible picture of electrons recombination into Cooper pairs, and  a  much weaker condition for the appearance of high-temperature superconductivity.
Another consequence is that the popular interpretation of Josephson qubits as macroscopic quantum systems is replaced by a picture of qubit states being superpositions of the ground state and the state containing only a single excited Cooper pair.
\end{abstract}
\maketitle

Fifty years after the discovery of superconductivity Bardeen, Schrieffer, and Cooper (BCS) presented a theory of this phenomenon which remained valid for the next fifty years \cite{BCS}. The BCS theory  contains a cut-off energy parameter determining the number of electrons participating in the pairing mechanism and initially suggested to be comparable to the superconducting gap. For many years this parameter remained unspecified \cite{Fey} but in the modern literature has been replaced by the Debye energy \cite{Tin} which is about hundred times higher than the gap. The relatively new experiment with flux qubit \cite{Wal} sheds a new light on that issue supporting the original hypothesis. The detailed analysis of the  microscopic Hamiltonians based on this hypothesis is presented in \cite{Ali} for all three types of Josephson qubits (charge, phase, and flux qubits \cite{Cla}). Here, only a simplified model of flux qubit is discussed.
\par
A flux qubit (FQ) is a small superconducting ring interrupted by  one or several Josephson junctions. The thickness of the ring is assumed to be not larger than the penetration depth for the magnetic field (typically $\sim 100 nm$), what allows to ignore spatial variations of the electric current density.  The Cooper pair's circular states $|k,\mu\rangle$, which produce current, are labeled by two quantum numbers. The first one $k= 1,2,..,K$  describes a Cooper pair in its center of motion reference frame and corresponds to a pair of time-reversed single electron states $|k_{\pm}\rangle$. Under standard half filling condition there are $K/2$ Cooper pairs in the sample at zero temperature. The second quantum number 
$\mu = 0, \pm 1, \pm 2,...$ accounts for the quantized circular motion of the Cooper pair along the ring. According to the Onsager hypothesis \cite{Ons} a single Cooper pair in a state $|k,\mu\rangle$ generates a quantized magnetic flux
$\mu \Phi_0$ where $\Phi_0 = h/2e$. Consider a configuration of Cooper pairs described by the occupation numbers
$n_{k\mu}$ which take values $0,1$, as Cooper pairs behave like \emph{hard core bosons}. The total magnetic flux is equal to $F\,\Phi_0$, where  $F=\sum_{k,\mu}\mu\, n_{k\mu}$, and the magnetic energy of such configuration is given by
\begin{equation}
\mathcal{E}_M = E_L \Bigl(\sum_{k,\mu}\mu\, n_{k\mu}\Bigr)^2 .
\label{magen} 
\end{equation}
where
\begin{equation}
E_L= K h^2/4m\ell^2,
\label{EL}
\end{equation}
$m$ is electron's mass and $\ell$ is the length of the loop. The formulas (\ref{magen}) and (\ref{EL}) are derived using the London's relation between superconducting current density and vector potential
$\mathbf j = - \frac{\mathcal{N}e^2}{m}\mathbf{A}$, the formula for the energy of a current in a magnetic field $\mathcal{E}= -\int \mathbf {j}\cdot\mathbf{A}\, d^3\mathbf{x} = \frac{K e^2}{m} A^2$ and the Onsager's magnetic flux quantization condition $\Phi = \oint\mathbf{A}\cdot d\mathbf{x}= \pm A\ell = F \Phi_0,$ $ F=0,\pm 1,\pm 2,..$.
Here, $\mathcal{N}$ is a density of superconducting electrons and $A$ denotes $|\mathbf{A}|$. Both quantities are assumed to be uniform in the sample what is true for thin enough loops (thickness $\leq 100 nm$). 
\par
For loops of micrometer diameters,  used to construct flux qubits, $E_L $ is much higher than the superconducting gap what means that the states with different fluxes $F\,\Phi_0$  are separated by large energy gaps in comparison with the other relevant energy scales for the system. To produce a qubit one has to switch on an external flux $\Phi_{ext}=\mu_{ext}\Phi_0$ which modifies the magnetic energy yielding
\begin{equation}
\mathcal{E}_M(\mu_{ext}) = E_L \bigl(F - \mu_{ext}\bigr)^2 .
\label{magen1} 
\end{equation}
Taking $\mu_{ext}\simeq 1/2$ one can reduce the magnetic energy difference $\delta\mathcal{E}_M = E_L |1- 2\mu_{ext}|$ between the unique ground state $|0\rangle$ with a flux $F=0$ and the excited levels  with $F=1$.
The stable states of a superconducting ring correspond to collective states of Cooper pairs with various flux population numbers  $N_{\mu}= \sum_{k} n_{k\mu}$ such that $\sum_{\mu} N_{\mu} = K/2$. In particular the ground state $|0\rangle$ has the total flux equal to zero with $N_{\mu}$ concentrated around $\mu=0$. The stable states with $F=1$  are obtained from the ground state by 
changing the populations $N_{\nu}\mapsto N_{\nu}-1 , N_{\nu+1}\mapsto N_{\nu+1}+1$ at the given $\nu$. Such states are denoted by $|0,\nu\rangle$. The presence of Josephson junction(s) does not produce transitions between $|0\rangle$ and $|0,\nu\rangle$ because those states have a global and collective character while the Josephson junction should be described by a localized Hamiltonian perturbation
$\hat{V}_{J}$. However, the transitions exist between $|0\rangle$ and the states $|0,\nu, k\rangle$ obtained from $|0,\nu\rangle$ by replacing one \emph{ground pair} with the indices $(\nu+1 ,k)$ by an \emph{excited pair}. Excited Cooper pairs appear already in the original BCS paper \cite{BCS}
and possess a clear interpretation in terms of permutational symmetry within a strong-coupling BCS model \cite{Thou}.
The corresponding transition amplitudes $\xi_{\nu k}= \langle 0,\nu,k|\hat{V}_{J}|0\rangle$ determine the state 
\begin{equation}
|1\rangle = \sum_{\nu, k}\tilde{\xi}_{\nu k}|0,\nu, k\rangle\ ,\ \tilde{\xi}_{\nu,k}=\frac{\xi_{\nu k}}{\sqrt{\sum_{\nu'k'}|\xi_{\nu' k'}|^2}}
\label{state1}
\end{equation}
which becomes separated from the others and together with the ground state $|0\rangle$ span the flux qubit Hilbert space.
\par
In the two level approximation the eigenstates of the qubit Hamiltonian are suitable superpositions  of $|0\rangle$ and $|1\rangle$ and the corresponding qubit frequency reads
\begin{equation}
f = \frac{1}{h}\sqrt{\bigl[ E_L(1- 2 \tilde{\mu}_{ext})\bigr]^2 + E_J^2} . 
\label{fFQ}
\end{equation}
Here, the Josephson energy $E_J= \sum_{\nu k}|\xi_{\nu k}|^2$ and  $\tilde{\mu}_{ext} ={\mu}_{ext}+ \delta$. The correction $\delta$  accounts for the fact that the excited pairs have higher energy than the ground ones (see \cite{Ali} for the details).
\par
The simple theory presented above can be now compared with the experiment of van der Wal et.al.\cite{Wal}.
The sample is an aluminum $5\mu m\times 5\mu m$ loop made of $450 nm$ wide and $80  nm$ thick lines. The Josephson junction loop is kept at the milikelvin temperatures and excited by the microwave radiation with frequencies in the range of $1-10\, GHz$. The measured dependence of the resonant frequency on the control parameter $\tilde{\mu}_{ext}$ in the range $0.495 \leq \tilde{\mu}_{ext}\leq 0.505$ can be used to test the formula (\ref{fFQ}). The off-set $\delta$ is observed also, but attributed to persistent currents. The linear dependence on $|1- 2 \tilde{\mu}_{ext}|$ far enough from $\tilde{\mu}_{ext} = 1/2$ is clearly visible as well as the level repulsion for $\tilde{\mu}_{ext}$ very close to $1/2$ (see FIG.1). 
%\fig 2
\begin{figure}[!ht]
\begin{center}
\includegraphics[width=0.8\textwidth]{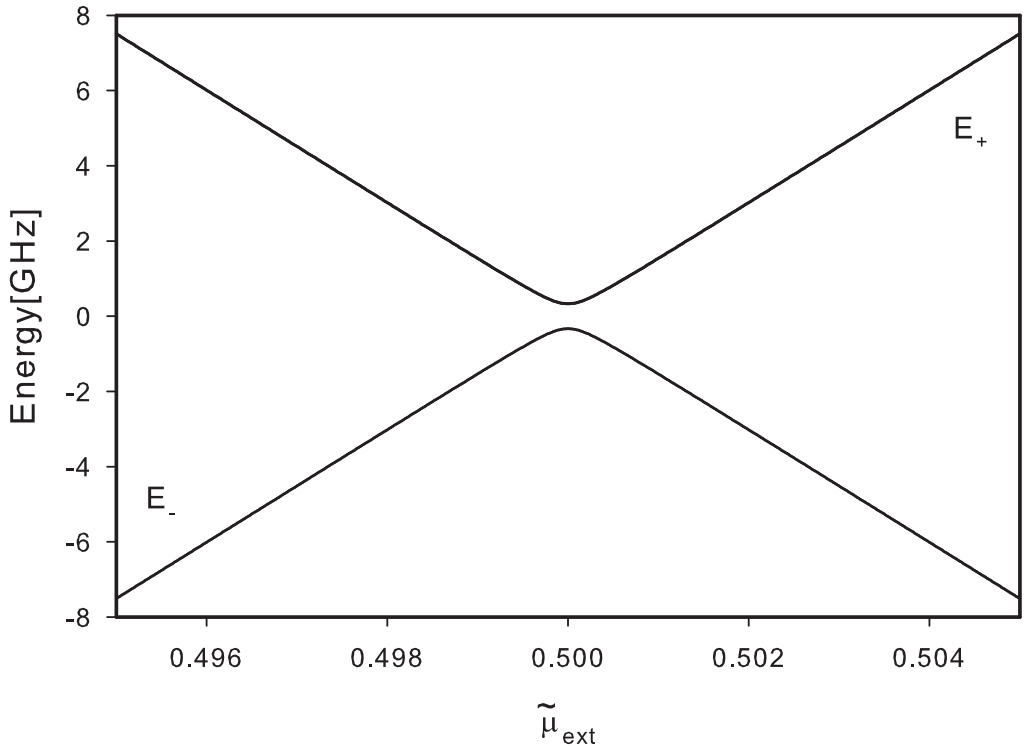}
\end{center}
\caption{Energy levels of the flux qubit (in GHz) as functions of the external magnetic flux (in $\Phi_0$ units) given by (\ref{fFQ}). A fit to the data of van der Wal et.al. with $E_L/h = 1500 GHz$ and $E_J/h = 0.66 GHz$.}
  \label{fig1}
\end{figure}
From those data the value $E_L/h \simeq 1.5 \times 10^3 GHz$ is extracted. Putting $\ell = 20 \mu m$ into (\ref{EL}) one obtains $K = 3.3\times 10^6$. The volume of the sample $\mathcal{V}= 0.72\times 10^{-12}\mathrm{cm}^3$ and hence, for the first time, one can compute directly from the experimental data the zero temperature density of Cooper pairs  in a superconductor (Al)
\begin{equation}
\kappa_{CP}[Al]= K/2\mathcal{V} = 2.5\times 10^{18}/\mathrm{cm}^3 .
\label{CPden}
\end{equation}
The obtained result can be discussed within the BCS theory \cite{BCS,Tin}. The basic predictions of the BCS concern the relations between the superconducting gap at zero temperature $\Delta(0)$,  the critical temperature $T_c$,
the coupling constant $g$ describing the magnitude of electron-phonon interaction \cite{g}, and  the cut-off 
parameter $\hbar\omega_{c}$
\begin{equation}
\Delta(0)= 1.76\, k_B T_c\ ,\  k_B T_c = 1.13\, \frac{\hbar\omega_{c}}{2\sinh{(\hbar\omega_{c}/g)}} .
\label{gap1}
\end{equation}
The energy cut-off $\hbar\omega_{c}$ limits the kinetic energy of the electrons participating in the BCS pairing mechanism to the interval $[E_F - \hbar\omega_{c}, E_F + \hbar\omega_{c}]$ ($E_F$ - Fermi energy) and determines the number $K$ of the corresponding electronic states by the expression \cite{Tin}
\begin{equation}
K =2\hbar\omega_{c}N(0).
\label{K}
\end{equation}
Here, $N(0)$ is the density of electronic Bloch states (excluding electron's spin) at the Fermi surface given by \cite{Kit}
\begin{equation}
N(0) = \mathcal{V}\frac{m}{2\pi^2 \hbar^2}\bigl(3\pi^2 \kappa_{el}\bigr)^{1/3}
\label{CP}
\end{equation}
where $\kappa_{el} $ is a density of electrons ($\kappa_{el}[Al] = 18.06\times 10^{22}/\mathrm{cm}^3$). After substitution one obtains  $\hbar \omega_{c}/2k_B = 1.3 K$ what is close to the critical temperature $T_c = 1.2 K$ for $Al$. It allows to formulate the following hypothesis concerning the parametrization of the BCS model:
\begin{equation}
\hbar\omega_{c} \simeq g \simeq \Delta(0) \simeq 2k_BT_c .
\label{hypo}
\end{equation}
Notice that (\ref{hypo}) agrees with the suggestion  $\hbar\omega_{c} \sim k_BT_c$ in the original BCS paper, as well as with the BCS relations (\ref{gap1}). On the other hand the assumption $\omega_{c}= \omega_{D}$ (Debye frequency) usually made in the modern literature implies
\begin{equation}
\hbar\omega_{c}= \hbar\omega_D >> g >>  \Delta(0) \simeq 2k_BT_c .
\label{hypomod}
\end{equation}
The choice $\hbar\omega_{c} \simeq g $ allows to replace the electron kinetic energy in the BCS Hamiltonian by a constant $E_F$ what leads to a much simpler exactly solvable model which predicts the relation $g=\Delta(0)= 2k_BT_c$ \cite{Thou}.
\par
Another consequence of (\ref{hypo}) is the following plausible picture of electrons recombination into Cooper pairs. For the temperatures $T > T_c$ only about $4k_BT N(0)$ \cite{el} electrons are thermally excited and  contribute to specific heat and conductivity; the rest is "frozen in the Dirac see". At the temperature $T_c$ those $4k_BT_c N(0)=2\hbar\omega_c  N(0)= K$ electrons begin to recombine such that at zero temperature all of them form $K/2$ Cooper pairs while the rest of electrons remain frozen all the time.

\par
The next consequence of (\ref{hypo}) is presented on
FIG.2 which shows the relations between the coupling constant $g$  and the critical temperature $T_c$ under the hypothesis (\ref{hypo}) and (\ref{hypomod}), respectively. 
%\fig 1
\begin{figure}[!ht]
\begin{center}
\includegraphics[width=0.8\textwidth]{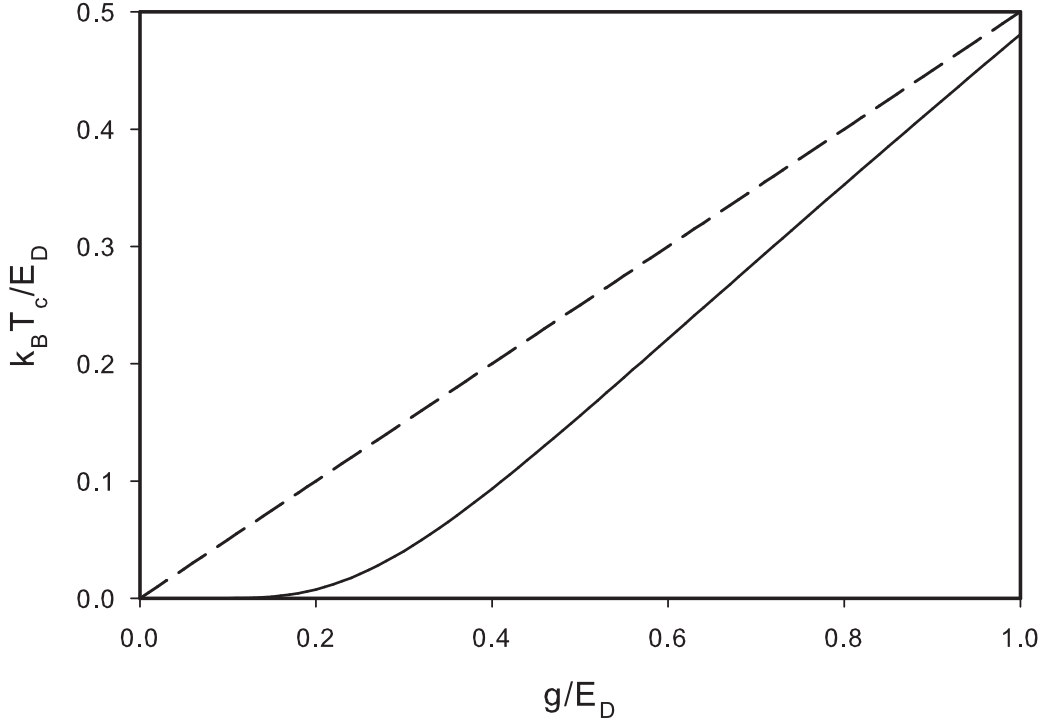}
\end{center}
\caption{Critical temperature as a function of the coupling constant (in Debye energy units $E_D =\hbar\omega_D$) for  $k_B T_c = 1.13\, {E_{D}}/2\sinh (E_{D}/g)$ (solid line) and  $k_B T_c = g/2$ (dashed line).}
  \label{fig2}
\end{figure}
Obviously, (\ref{hypo}) allows  to obtain quite high critical temperatures at the cost of moderate increase of the electron-phonon coupling, while (\ref{hypomod}) strongly suppresses the increase of $T_c$. For example, to obtain the critical temperatures $1.2 K (Al)$ and $9.8 K(Nb)$ one needs, assuming (\ref{hypomod}), the coupling constant $g/k_B$ equal to $72 K$ and $80 K$, respectively. The same values of $g$, assuming (\ref{hypo}), yield  critical temperatures $36 K$ and $40 K$. Therefore, (\ref{hypo}) suggests that the  phonon-mediated pairing mechanism could be sufficient to explain high-temperature superconductivity, at least for a certain class of such materials \cite{Bon}.
\par
In conclusion, it has been shown that the experimental results for  a small Josephson junction loop  interpreted in the light of the presented model do not only support the original BCS parametrization, what can be relevant for the theory of high-temperature superconductivity, but also change the physical picture of a flux qubit itself. Instead of the standard interpretation in terms of two macroscopic quantum states corresponding to the motion of millions of Cooper pairs, here the qubit's basis consists of the ground state and the state which differs from the ground one by a single excited pair, only.  Obviously, more experimental evidence concerning flux qubits of different sizes and made of different materials is needed as well as  the first principle theoretical justification of the hypothesis (\ref{hypo}).

\textbf{ Acknowledgments} The author thanks W. Miklaszewski for the assistance. This work is supported by the Polish Ministry of Science and Higher Education, grant PB/2082/B/H03/2010/38.


\begin{thebibliography}{99}
\bibitem{g}
The coupling constant $g$ is equal to the product $\hbar\omega_c\,V\,N(0)$ in the standard notation used in \cite{BCS,Tin}.
\bibitem{el}
This simple estimation roughly agrees with the magnitude of the electronic specific heat \cite{Tin}. The factor 2 accounts for an electron's spin and $2k_BT$ is the thickness of the energy shell around the Fermi level affected by thermal fluctuations.
\end{thebibliography}
\end{document}